# Predictive Modeling of non-viral Gene Transfer

Gerlinde Schwake[3,§], Simon Youssef[1,3 §], Jan-Timm Kuhr[1,2,3,§], Sebastian Gude[3], Maria Pamela David[3], Eduardo Mendoza[1,3,4], Erwin Frey[1,2,3] and Joachim O. Rädler[1,3 #]

[1] Center for NanoScience (CeNS) and [2]Arnold Sommerfeld Center for Theoretical Physics,

[3]Fakultät für Physik, Ludwig-Maximilians-Universität, Geschwister-Scholl-Platz 1, D-80539 München, Germany

[4] Institute of Mathematics, University of the Philippines, Diliman, Quezon City 1101

§ G.S., S.Y. and J-T.K. contributed equally to this work.

[#] To whom correspondence should be addressed.

E-mail: Gerlinde.Schwake@physik.lmu.de

Manuscript information:

Word and character counts:

5649 and 32902 resp.

# Abstract

In non-viral gene delivery, the variance of transgenic expression stems from the low number of plasmids successfully transferred. Here, we experimentally determine Lipofectamine- and PEI-mediated exogenous gene expression distributions from single cell time-lapse analysis. Broad Poisson-like distributions of steady state expression are observed for both transfection agents, when used with synchronized cell lines. At the same time, co-transfection analysis with YFP- and CFP-coding plasmids shows that multiple plasmids are simultaneously expressed, suggesting that plasmids are delivered in correlated units (complexes). We present a mathematical model of transfection, where a stochastic, two-step process is assumed, with the first being the low-probability entry step of complexes into the nucleus, followed by the subsequent release and activation of a small number of plasmids from a delivered complex. This conceptually simple model consistently predicts the observed fraction of transfected cells, the cotransfection ratio and the expression level distribution. It yields the number of efficient plasmids per complex and elucidates the origin of the associated noise, consequently providing a platform for evaluating and improving non-viral vectors.

# Introduction

Non-viral gene delivery systems have evolved over the last decade into widely-used vectors for exogenous DNA delivery to eukaryotic cells. Synthetic cationic lipids and polymers, in particular, are used in molecular biology for transgene expression, and are being further refined for use in DNA-based therapies (Ferber 2001; Roth and Sundaram 2004) (Patil et al. 2005). Despite considerable progress in the efficiency and characterization of vectors, important aspects of the delivery pathway and transfer kinetics remain poorly understood, including how artificial vectors are taken up, transported to the nucleus, and how these factors collectively influence the expression characteristics of a cell population. Current understanding from intracellular studies of transgene delivery includes the following steps: DNA-vector complex uptake via the endosomal pathway, followed by endosomal escape and cytoplasmic transport, nuclear entry, vector unpacking and transcription initiation (Roth and Sundaram 2004) (de Bruin et al. 2007; Kircheis et al. 2001; Lechardeur et al. 2005; Safinya 2001; Suh et al. 2003). These processes are accompanied by a huge loss of material and temporal delays. It is therefore not surprising that transfected cells in a culture respond very heterogeneously over time, notably in terms of the expression onset time ($t_{on}$) and the maximum expression levels attained. It is generally accepted that the expression behavior of a single transfected cell is stochastic, yet cell culture averaged expression levels are reliable indicators of gene transfer efficiency.

Flow cytometry is commonly used to measure fluorescence distributions over a population at a rate of up to 10000 cells per second (Longo and Hasty 2006). High-content single cell assays, in contrast, are particularly suitable for investigating the dynamics and heterogeneity of clonal cell populations, since individual cells can be followed with a high temporal resolution. In addition, quantitative image analysis has been successfully improved to reliably convert fluorescence intensities into copies of molecules, hence paving the path to follow 'gene expression by numbers' (Rosenfeld et al. 2005).

In this paper, we analyze gene expression following non-viral gene delivery, with focus on the variance of expression levels. The expression of genes exhibits all-or-nothing characteristics (Hume 2000) and additional stochasticity exists in transcriptional regulation (McAdams and Arkin 1999) (Rao et al. 2002). Elowitz et al. have analyzed noise in bacterial gene expression and elucidated the distinction between extrinsic and intrinsic noise, i.e. the contribution of fluctuations in cellular components and inherent stochasticity of the biochemical processes during gene expression (Elowitz et al. 2002). The extrinsic variance of gene expression within a clonal population of eukaryotic cells has been investigated in the light of stochastic theories(McAdams and Arkin 1997; Volfson et al. 2006) (Blake et al. 2003) (Raser and O'Shea 2004). It was only recently that attempts were made to generate models for transgene expression following non-viral gene delivery (Varga et al. 2000) (Varga et al. 2001) (Dinh et al. 2007) (Zhou et al. 2007). Computational modeling might greatly enhance our understanding of gene transfer and aid in elucidating the nature of the underlying transport barriers. Many of the issues regarding cell entry and intracellular transport are shared with attempts to model viral infection (Varga et al. 2005) (Douglas 2008).

In this article, we used quantitative single cell time-lapse microscopy combined with mathematical modeling to analyze the variability in transgene expression (Fig. 1). From the synthetic delivery agents currently being evaluated for therapeutic use, we chose polyethyleneimine (PEI) (Boussif et al. 1995) and the commercial Lipofectamine 2000, as cationic polymer and lipid model systems, respectively. Both synthetic vectors are able to condense plasmid DNA into DNA-nano particles, denoted as cationic lipid- (cationic polymer-) DNA complexes or just "complexes". Distributions of the expression onset times and expression steady state levels were evaluated for both vectors.

Data are well described by a stochastic delivery model, which is based on the assumption that in a decisive step, only a small number of complexes enter the nucleus through a stochastic process. Out of these complexes, only a fraction of the plasmid load is expressed (Fig. 1). The theoretical model is further corroborated by a cotransfection analysis, i.e. the case of the simultaneous transfection using two distinguishable plasmids encoding for CFP and YFP. It is shown that this model consistently describes the fraction of transfected cells and the observed expression level distribution. As a consequence the effective size of a stochastically delivered unit of plasmids (complex) can be determined.

## Materials and Methods

**Cell culture.** A human bronchial epithelial cell line (BEAS-2B, ATCC) was grown in Earle's MEM supplemented with 10% FBS at 37°C in a humidified atmosphere, 5% $CO_2$ level. Transfection was performed on both non-synchronized and synchronized cultures. A thymidine kinase double-block was performed to synchronize cells.

**Transfection.** BEAS-2B cells were grown to 80% confluence from an initial seeding density of 1 x $10^5$ cells/well in six-well plates 24 hours before transfection. Cells were washed and the medium is replaced with 1 ml OptiMEM/well immediately before transfection. Optimized transfection procedures were performed using either 2% v/v Lipofectamine$^{TM}$2000/OptiMEM or PEI (N/P=8)/HBS; 1 µg of pEGFP-N1 or pd2EGFP-N1, is used for transfecting each batch of cells. The transfection medium was prepared either by adding the Lipofectamine or the PEI solution to the plasmid solution. After the transfection media were allowed to stand for 20 minutes the cells were incubated with 200 µl/well Lipofectamine or PEI transfection medium for 3 hours at 37°C, 5% $CO_2$ level. After 3 hours of incubation the medium was removed, and cells were washed with PBS. Cells were reincubated with Leibovitz's L-15 Medium with 10% FBS prior to EGFP expression monitoring.

**Cotransfection.** Cotransfection was performed with two kinds of preparations containing the same molar amount per plasmid. For one preparation, ECFP/Lipofectamine, EYFP/Lipofectamine, ECFP/PEI and EYFP/PEI were complexed separately. For the other, a mixture of ECFP and EYFP hetero-complexes were complexed with Lipofectamine or PEI. Transfection using either the hetero-complexes (pre-mixed) or a mixture of homo-complexes (post-mixed) was performed as previously described. Cells were reincubated in growth medium and CFP and YFP expression was monitored by fluorescence microscopy after 24 hours.

**Data acquisition and quantitative image analysis.** Images were taken at 10× magnification, with a constant exposure time of 1s, at 10 minute-intervals for at least 30 hours post-transfection. Fluorescence images are consolidated into single image sequence files. Negative control images were taken to assess lamp threshold values and autofluorescence, and were subtracted from corresponding image sequence files to eliminate autofluorescence effects. To capture cell fluorescence over the entire sequence, regions of interest (ROIs) were manually defined around each cell (Fig. 2). Changes in total gray measurements in individual ROIs were determined for each time point.

# Results

**Time lapse microscopy and single cell EGFP expression**

A cell line of lung epithelium cells was transfected with a plasmid encoding for the green fluorescent protein (EGFP). Transfection protocols for PEI- and Lipofectamine-mediated delivery followed standard procedures and are described in detail in the supplementary online information. We denote the time of the gene vector administration to the cell culture as $t$ = 0h. Transfection medium was removed and cell growth medium added at t = 3h. Single-cell EGFP expression was monitored by automatically taking sequences of fluorescence micrographs from 25 view fields at 10-minute intervals. Fig. 2a shows a representative sequence from a Lipofectamine transfection experiment, with the initial bright field image, as well as the EGFP fluorescence at $t$ = 4, 8 and 12 hours post-transfection. These images demonstrate heterogeneity in both the expression onset times and levels of exogenous gene expression. It is observed that the number of fluorescent cells increases with time; at the late stage (~30h), the ratio of fluorescent to non-fluorescent cells is about 23% and 30% for PEI- and Lipofectamine-mediated transfection, respectively. A total of 500-1500 cells were monitored in parallel within one time-lapse experiment. Individual time courses of the total fluorescence per cell were evaluated by image processing from data stacks as shown in Fig. 2b and described in the supplementary online information. Fig. 2c shows a series of representative time traces from one transfection experiment, illustrating the significant variance in both the expression onset time and EGFP expression level. The typical sigmoidal shape of the time courses is well described by the phenomenological function,

$$I(t) = \frac{I_{max}}{2}\left[1 + \tanh\left((t - t_{1/2})/t_{rise}\right)\right], \qquad [1]$$

which allows the determination of the maximal fluorescence plateau value ($I_{max}$) the time of half-maximum ($t_{1/2}$) and the characteristic rise time, $t_{rise}$. The fluorescence intensities were converted into molecular units using EGFP standard beads for calibration (see supplementary online information). In the remainder of the text we will give the fluorescence intensity $I_{max}$ in units of EGFP numbers $G$. Eq. **1** proved to be robust for automated data analysis, facilitating the accumulation of statistics for a large number of individual cells. In order to determine the time points of expression onset, $t_{on}$, we use $t_{1/2}$ - $t_{rise}$ as an approximation due to the lack of a well-defined point of onset as shown in Fig. 2d. Since the timing of gene expression is expected to be dependent on the cell cycle, we investigated the distribution of $t_{on}$ for synchronized and non-synchronized cells. To this end, cells were arrested at the G1/S-Phase transition using a thymidine kinase double-block (Merrill 1998).

**Distribution of expression onset times**

Fig. 3 summarizes the measured distribution functions of $t_{on}$ and $I_{max}$ for GFP expression after transfection of non-synchronized and synchronized cultures with PEI- and Lipofectamine-based complexes respectively. Expression onset times range from 5 and 25 hours for both PEI- and Lipofectamine, indicating the existence of a time window during which plasmids are successfully transcribed in the nucleus. The distribution for Lipofectamine clearly peaks at an earlier time (~8h) compared to PEI, (~16h). On cell cycle synchronization, the distribution of onset times sharpen and become peaked at about $t = 15h$ for both PEI and Lipofectamine. Bright field images reveal that most synchronized cells divide 12 hours after transfection. This is consistent with the fact that transfection was carried out in mid-S-phase, three hours after release from the thymidine kinase double-block. This shows that plasmid activation occurs about three hours after the M-phase. Furthermore, since cell cycle synchronization suppresses expression at earlier times, there is evidence that the delivery process depends on the cell cycle dependent breakdown of the nuclear membrane. This is consistent with previous studies claiming that mitosis enhances transgene nuclear translocation in cationic lipid gene delivery (Tseng et al. 1999) (Mortimer et al. 1999) (Brunner et al. 2000). We also find that synchronization leads to a 2-fold higher steady state expression for PEI- (from $2.6 \cdot 10^6$ to $4.3 \cdot 10^6$ average EGFP molecules per cell) and Lipofectamine- (from $2.9 \cdot 10^6$ to $5.1 \cdot 10^6$ average EGFP molecules per cell) mediated transfection, consistent with earlier observations on ensemble averaged data (Brunner et al. 2000).

**Modeling steady state gene expression**

In order to analyze the distribution of expression steady states, we introduce a mathematical model that describes EGFP expression after transfer and nuclear translocation of complexes containing exogenous EGFP plasmids. Stochasticity due to nuclear translocation of the plasmid complexes and the intra-nuclear activation will give rise to a probability distribution $P(X)$ for $X$ successfully expressed plasmids (see Fig. 4a). In the first stage of analysis, we describe the expression of EGFP from a single activated plasmid in a linear deterministic model and neglect any cell-cell variability. Based on the biochemical reactions shown in Fig. 4b we denote the ensuing rate equations:

$$\dot{R} = s_A X - \delta_R R \qquad [2]$$

$$\dot{U} = s_P R - (k_M + \delta_U) U \qquad [3]$$

$$\dot{G} = k_M U - \delta_G G \qquad [4]$$

Here, $R$ denotes the number of RNA molecules, $U$ the number of unfolded polypeptide chains, and $G$ the number of folded EGFP proteins. $s_A$, $s_P$ and $k_M$, denote the rate constants for transcription, translation and EGFP maturation, $\delta_R$, $\delta_U$, and $\delta_G$ denote the degradation constants of each product, respectively. The degradation rates of folded ($\delta_G$) and unfolded protein ($\delta_U$) are assumed to be equal, since the same proteases are involved (Leveau and Lindow 2001). A plasmid degradation term was omitted, since its occurrence is predicted to be negligible within the time frame considered (Subramanian and Srienc 1996). Literature values for the individual kinetic rates are summarized in Table I. Equations **2-4** can be solved analytically. For the steady state value, a linear relation

$$I_{max} = G(t \to \infty) = \frac{k_M s_P s_A}{\delta_G \delta_R (k_M + \delta_G)} \cdot X \qquad [5]$$

between the expression level $I_{max}$ = [GFP] and the number of expressed plasmids $X$ = [plasmids] is obtained:

$$[GFP] = k_{exp} \cdot [plasmids] . \qquad [6]$$

Here, $k_{exp}$ denotes an effective expression factor, corresponding to the number of proteins expressed per transcribed plasmid in the steady state. With the values given in Table I, we find $k_{exp} \approx 4 \cdot 10^6$ molecules/plasmid, which compared to the experimental number of molecules ($1\text{-}15 \cdot 10^6$), results in the remarkable finding that the number of plasmids $X$ is of order one. This implies that most of the variance in expression level originates from stochastic variations in the small number of plasmids, such that the distribution of GFP expression is determined by the distribution of successfully delivered plasmids, $P(G) \sim P(X)$.

To further substantiate this conclusion, we designed an experiment where the expression factor $k_{exp}$ is deliberately modified through the use of destabilized EGFP. It has a 14-fold higher degradation rate ($\delta_{desG}$) due to an additional amino acid sequence (PEST), which makes it more susceptible to proteolysis (Kain 1999). Figs. 3e and 3f display the shift in the steady state distribution of $I_{max}$, shown in a logarithmic scale. As predicted above, the shape of the distribution function is almost unchanged for both PEI- and Lipofectamine-mediated transfection. In addition, the peak positions shifted by a factor 12.5, which is close to the value 14.3 predicted from Eq. **5**.

**Modeling transfection noise**

Unlike in chromosomal DNA, which contains a fixed number of genes, the transfection experiments discussed here result in the delivery of a variable number of genes per vector. We model gene delivery as a two-step stochastic process as shown in Fig. 4a. As we will argue in the following, a two-step model is the simplest model that is in accordance with the experimental data. The model consists of (i) the nuclear translocation with probability $\mu$ of complexes containing an average of m plasmids and (ii) intra-nuclear activation of plasmids, with probability $q$. Whereat the probability q subsumes all phenomena promoting or interfering with transcription such as DNA methylation or complexation. We assume that the first process, the delivery of complexes to the nucleus, is rare and statistically independent, yielding a Poisson distribution for the number of delivered complexes C:

$$P(C) = \frac{\mu^C}{C!} e^{-\mu} \qquad [7]$$

characterized by its mean value $\mu$. Secondly, the independent activation of a plasmid in the nucleus is described by a Bernoulli process with success probability q. The concatenation of both processes results in an expression for *P(X)* which retains the characteristics of a Poissonian. Mathematical details of its derivation can be found in the supplementary data. Fig. 5 shows the calculated distribution of activated plasmids, *P(X)* (red bars), to the measured experimental protein distribution, $P_{exp}(G)$ (green bars). In addition a theoretical protein distribution is shown as black lines. $P_{theo}(G)$ is obtained from *P(X)* by additionally accounting for noise in gene expression, where we have used a relative magnitude of 0.3 for post-transfectional noise from the literature (see supplementary data). Note that the x-axis of the distributions P(G) are rescaled by the factor $k_{exp}$ according to Eq.6. The agreement between experiment and model is remarkable considering that there is only one free parameter in the fit. This is due to the fact that two additional experimental constraints have to be met. These are the measured fraction of transfected cells, TR, defined as the percentage of cells expressing one or more plasmids, and the average number of GFP molecules per cell, <G>, determined by calibration. The parameters $\mu$ and m·q are fixed by these constraints. The remaining unknown is the expression factor, $k_{exp}$, which is determined by the fit shown in Fig. 5. We obtain $k_{exp} \approx 1 \cdot 10^6$, $m_{eff} \approx 3$, and $\mu \approx 0.3\text{-}0.5$.

The red curve in Fig. 5 represents the distribution of successfully expressed plasmids. This distribution, which is directly related to the number of expressed GFP protein though Eq. 6, has a well defined mean value given by:

$$\langle [plasmids] \rangle = \mu \cdot m \cdot q \qquad [8]$$

The transfection ratio, *TR,* is related to this mean plasmid number. It depends on the average number of complexes delivered μ and the effective probability $\tilde{q}$ that from any given complex at least one plasmid is transcribed (we present a complete derivation of these quantities in the supplementary data).

$$TR(\mu, m, q) = 1 - \exp\{-\mu \cdot \tilde{q}\} \qquad [9]$$

For the data shown in Fig. 6, TR is of order 20%.

**Cotransfection and correlated delivery**

One important ingredient of our model is the delivery of DNA in units or complexes and the subsequent correlated coexpression of multiple plasmid copies. This assumption is closely related to the question of whether DNA-complexes fully dissociate before nuclear entry or complexes enter the nucleus as a whole. To elucidate this issue, we studied cotransfection of two distinguishable plasmids (CFP and YFP) and analyzed the outcome of transfection using pre-mixed and post–mixed complexes. Pre-mixed complexes contain both CFP- and YFP-plasmids in a single complex, whereas post-mixed complexes contain either CFP- or YFP-plasmids (for clarity see Figs. 6a and 6b). The steady-state CFP/YFP expression was analyzed at 24h post-transfection. We define the cotransfection ratio, *r*, as the number of cells expressing both CFP and YFP divided by the number of cells expressing either CFP or YFP. We find that the cotransfection ratio increases from 12.9% for post-mixed complexes to 21.9% for premixed complexes. The significant difference could not be explained, if complexes were completely dissolved in the cytosol and delivery of plasmids was independent from the complexes. The two-step delivery model, however, naturally explains the discrepancy between pre-mixed and post-mixed complexes. Based on our model an analytical expression for the cotransfection ratio can be derived (see supplementary data) which predicts correctly the measured cotransfection ratios, if the same parameters are used as determined from the EGFP distribution function.

## Discussion

We have measured the distribution of expression onset times and steady-state expression levels derived from single cell fluorescence time courses. Distributions of onset times of PEI and Lipofectamine collapse on a single curve for synchronized cell cultures, suggesting a universal cell cycle-dependent gene delivery mechanism. Synchronized cells exhibit a broad Poissonian distribution in expression levels and cotransfection experiments reveal correlations in the delivery probability for plasmids contained in one complex. Invoking Occam's razor, we analyzed the findings in terms of an idealized minimalist model of gene transfection, which describes gene delivery as a two-step stochastic process. Yet our model proves to have considerable predictive power by relating measurable quantities such as the overall transfection efficiency, the cotransfection probability and the shape of the gene expression distribution with each other. Thus, the model allows to derive the expression factor, the number of activated plasmids per complex and the average number of delivered complexes from the measured single cell transfection statistics. The model also elucidates the origin of expression variance, separating the noise due to small number fluctuations of complexes, which is inherent to the delivery process and extrinsic sources of noise due to cell-cell variability.

In our gene expression model we refer to complexes as units of coherently delivered plasmids. Those indirectly infered complexes are consistent with but not necessarily identical to the complexes described in many physico-chemical studies of PEI and lipofectamine mediated transfection. Cationic-lipid complexes are known to form multi-lamellar aggregates that contain a large number of plasmids (Zabner et al. 1995) (Rädler et al. 1997) (Lasic et al. 1997). Following endocytotic uptake and release, the complexes slowly dissociate in a stepwise, unwrapping mechanism (Lin et al. 2003) (Kamiya et al. 2002). PEI complexes are torroids or rods with a typical hydrodynamic radius of 100nm (Boussif et al. 1995) (DeRouchey et al. 2005), they have been seen to be actively transported inside cells (de Bruin et al. 2007) and to accumulate in the periphery of the nucleus (Suh et al. 2003). Both scenarios describe a situation where numerous small complexes have equal chances of entering the nucleus during the course of mitosis, which is consistent with our model assumptions. Microscopy studies have argued favorably for complexes being at least not fully dissolved at the final delivery stage (Lin et al. 2003) (Tseng et al. 1999). However, single nuclear entry events have not been documented explicitly. The probability of transgene expression in the nucleus again depends on the nature of the transfection agent. Pollard et al reported that cationic lipids but not PEI prevent gene expression when complexes are directly injected in the nucleus(Pollard et al. 1998). Such findings can only be consolidated with our model, if the delivered complexes transform during the course of the delivery, rather than being the same physical complexes as originally prepared under in vitro conditions. Within the context of our model we restrict ourselves to a narrowed meaning of "complexes" as units of plasmids that are co-delivered. In this framework, we determine the average number of successfully delivered complexes and the effective number of activated plasmids per complex from the analysis of single cell statistics. It will be interesting to corroborate the physical fate and the expression outcome of single complexes by high resolution studies in single cells (de Bruin et al. 2007).

The method to use transfection assays based on automated high-throughput microscopy combined with image processing might evolve into a routine tool for the assessment of transfection efficiency. In contrast to ensemble averaged fluorescence or luminescence data, single cell assays yield precise distribution functions and single cell expression dynamics, which allow a more detailed comparison to theoretical models. As shown here, the analysis of steady state expression levels provides access to the probability of successful plasmid delivery ($P(X)$) and yields an absolute number for the expression factor ($k_{exp}$). In forthcoming work we will discuss in more detail the distribution of expression onset times and the expression dynamics. We expect that our particular mathematical model can be adapted to a wider class of transfection agents and different types of cells. Their distinct transfection ratios, rate constants and numbers of effective complexes will become even more meaningful in the context of comparative theoretical modeling. A combined experimental and modeling approach will hence help to identify rate-limiting barriers to gene transfer and will result in improved data comparability, making it a versatile tool in the continuous evaluation and improvement of existing synthetic vectors.

**Additional information** regarding the transfection assays, image processing, data analysis and the mathematical model are available under **supplementary data**.

## Acknowledgements

We thank Josef Rosenecker, Carsten Rudolph and Ernst Wagner for fruitful discussions and Susanne Kempter for continuous support. The work was supported by the Deutsche Forschungsgemeinschaft through grant SFB486-B10 and SFB TR12, through the Excellence Cluster "Nanosystems Initiative Munich (NIM)", and through the LMU innovative program "Analysis and Modelling of Complex Systems". S.Y., J-T.K. and M.P.D. gratefully acknowledge scholarships from Microsoft Research, the IDK NanoBioTechnology and the DAAD, respectively.

**Table Legends**

**Table I.** Literature values for the kinetic rates of the linear gene expression model

**Figure Legends**

**Figure 1**. Experimental setup for single cell transfection experiments (upper part) and key elements of the theoretical model (lower part). EGFP-encoding plasmids and cationic agents form complexes, which are administered to eukaryotic cell cultures. Automated single cell microscopy yields statistics on phenotypic expression of EGFP. For the delivery of plasmids to the nucleus, stochastic effects are important, while the following expression of fluorescent proteins can be described in a deterministic fashion.

**Figure 2**. Acquisition of single cell time series. (**a**) Microscopy viewfields from a Lipofectamine transfection experiment. The first frame is a bright field (BF) control image. Fluorescence image sequences are taken automatically at ten minute intervals for at least 30 hours. (**b**) Definition of regions of interest (ROIs), total gray value measurement and conversion to the number of EGFP molecules. (**c**) Representative time-courses of EGFP expression in individual cells following PEI-transfection. The population average (red) is plotted to demonstrate its linear increase in contrast to the sigmoidal shape of the individual traces. (**d**) Characteristic parameters of expression are obtained by fitting the heuristic function **1** (red) to the recorded fluorescence time course (black). The time of expression onset, $t_{on}$, is calculated from the time of half-maximal expression $t_{1/2}$ and the slope at that point.

**Figure 3**. EGFP expression statistics for PEI- and Lipofectamine-mediated transfection. Distributions of expression onset times $t_{on}$ (**a,b**) and maximal expression values $I_{max}$ (**c,d**), for PEI- (red) and Lipofectamine- (dashed black) mediated transfection depict strong variability within the cell cultures. The total number of expressing cells was 23% out of 560 for PEI and 30% out of 502 in the case of Lipofectamine. (**b,d**) thymidine kinase - synchronized cultures with 40% out of 1981 and 30% out of 1797 cells expressing EGFP for PEI and Lipofectamine, respectively. For synchronized cells, expression onset time distributions coincide for Lipofectamine and PEI, indicating that transfection is more likely to happen in specific phases of the cell cycle. Distributions for $I_{max}$ (given in units of EGFP molecules) cannot be explained by post-transfectional sources of fluctuations alone. (**e, f**) Effect of the altered expression rates on the distribution of maximal expression levels $I_{max}$. Distributions for d2EGFP (gray) and EGFP (red) transfected with Lipofectamine (**e**) or PEI (**f**) are shown. d2EGFP, which has a higher degradation rate, exhibits a systematic shift of the $I_{max}$ distribution compared to EGFP, independent of the vector used. Besides this shift, a change in the number of proteins per active plasmid, $k_{exp}$, preserves the shape of the distribution. This suggests that the shape is determined during plasmid delivery prior to expression.

**Figure 4**. Theoretical model for transfection and gene expression. (**a**) Our model of plasmid delivery consists of several stochastic components. The number of complexes $C$ delivered per cell is Poisson-distributed, with mean $\mu$. Each complex carries a random number of plasmids, described by a Poisson distribution with mean $m$. Finally, each plasmid has an activation probability $q$, resulting in a Binomial distribution of active plasmids $X$ out of the total number of delivered plasmids. With this approach, the overall distribution, $P(X)$, of actively expressing plasmids can be derived. (**b**) Deterministic model of EGFP expression including transcription ($s_A$), translation ($s_P$) and protein maturation

($k_M$). mRNA ($R$), unfolded proteins ($U$) and GFP ($G$) are degraded with rates $\delta_R$, $\delta_U$ and $\delta_G$, respectively. Solving the corresponding rate equations, the steady state distribution of fluorescent proteins, $P(G)$, can be related to that of active plasmids, $P(X)$.

**Figure 5**. Comparison of single-cell data with the theoretical model. The theoretical EGFP distribution (black) is intimately connected with the underlying distribution of expressing plasmids (red). To facilitate comparison, the protein distribution has been scaled down by the average number of proteins per active plasmid in steady state, $k_{exp}$. (**a,b**) For synchronized cultures the measured protein distribution (green) is fitted very well by our theoretical model (black). The fit for PEI transfection (**a**) yields an average number of delivered complexes, $\mu = 0.53$, and an average number of activated plasmids per complex, $m_{eff} = 3.2$. In the case of Lipofectamine (**b**), we find $\mu = 0.37$ and $m_{eff} = 3.2$.

**Figure 6**. Correlated delivery in CFP/YFP cotransfection with post-mixed (uni-colored) complexes (left) and pre-mixed (dual-colored) complexes (right). (**a,b**) Post-mixed (uni-colored) and pre-mixed (dual-colored) complexes carry different plasmid content, but take the same pathway to the nucleus. (**c,d**) Superposition of CFP and YFP fluorescence after transfection reveals a qualitatively different expression pattern for the two distinct experimental protocols. Cyan fluorescence is slightly displaced to permit identification of cotransfected cells. All micrographs are artificially colored.

Figure 1

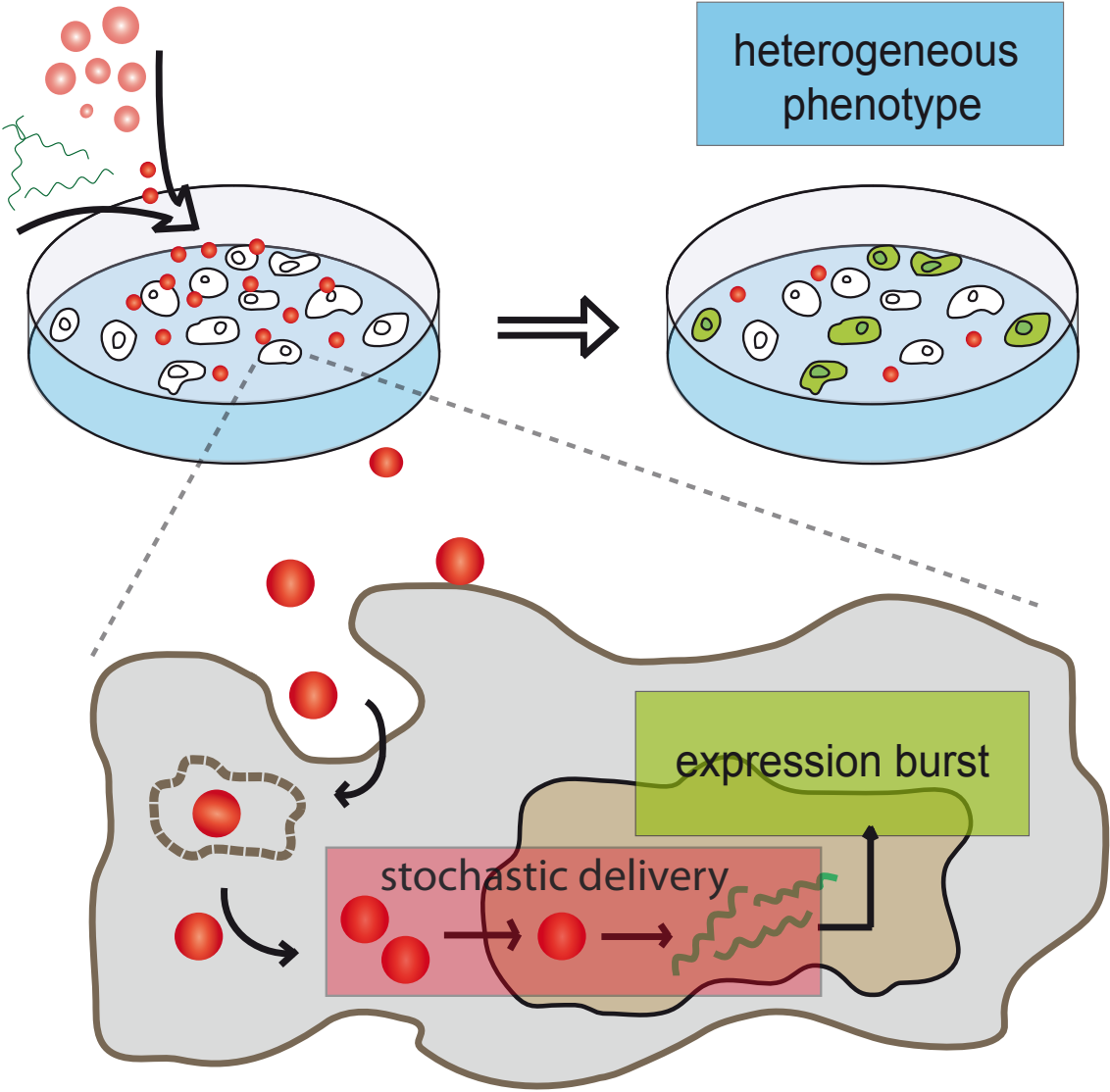

Figure 2

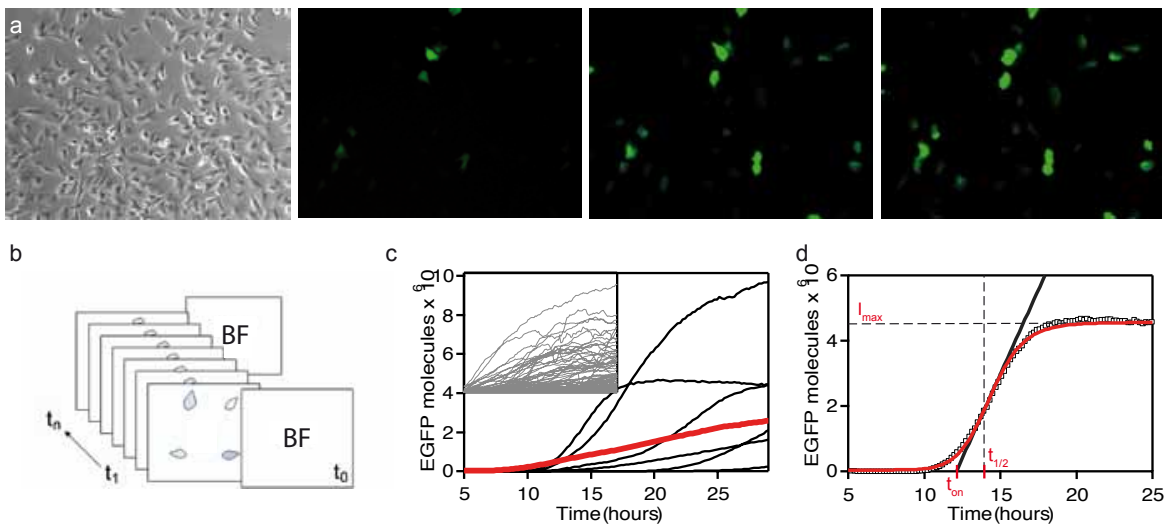

Figure 3

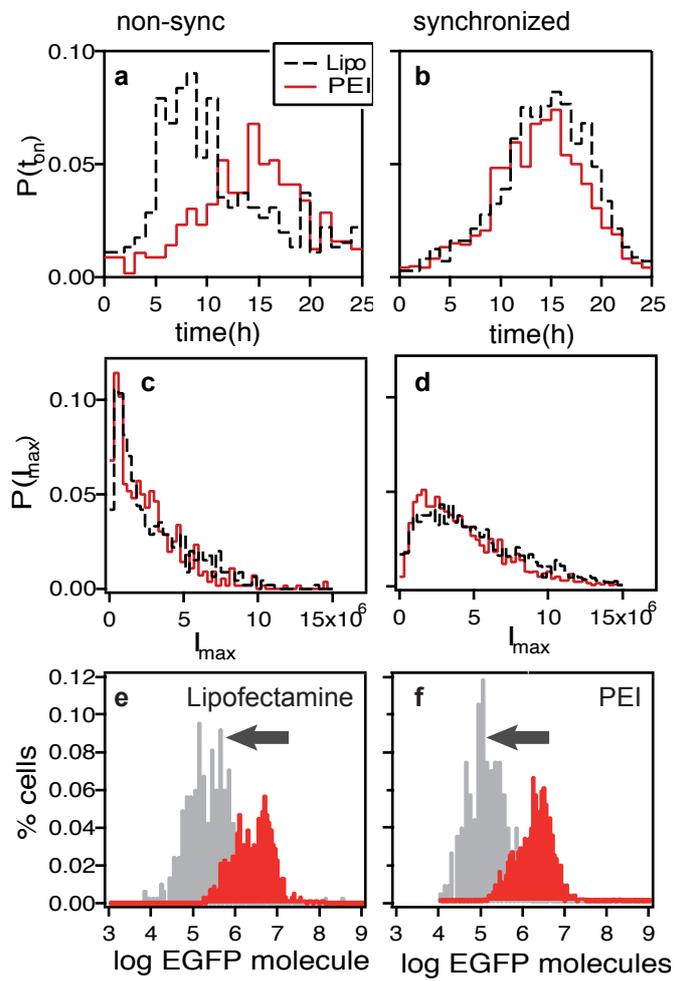

Figure 4

a

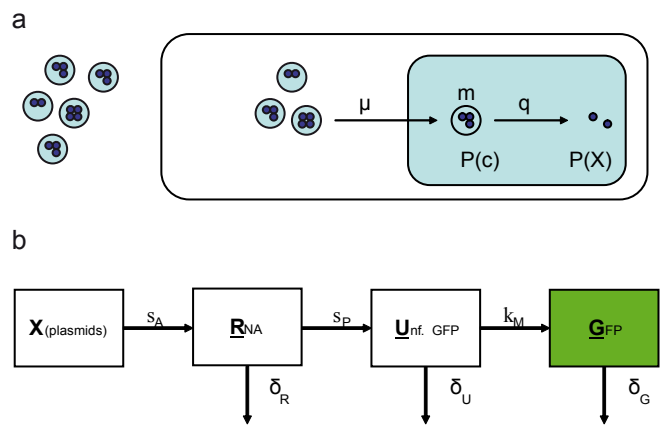

b

```
X(plasmids) --s_A--> R_NA --s_P--> U_nf. GFP --k_M--> G_FP
                 |δ_R          |δ_U              |δ_G
```

Figure 5

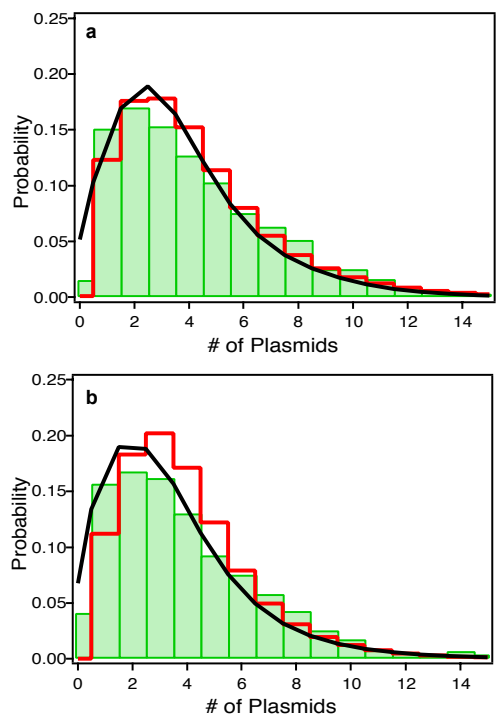

Figure 6

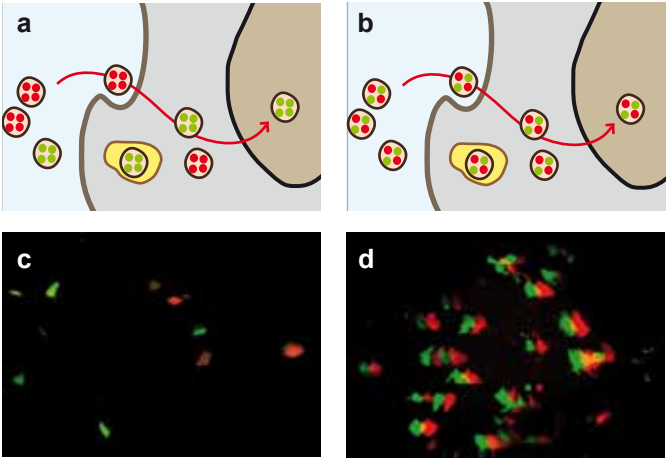

# Predictive Modeling of non-viral Gene Transfer

Gerlinde Schwake, Simon Youssef, Jan-Timm Kuhr, Sebastian Gude, Maria Pamela David, Eduardo Mendoza, Erwin Frey, Joachim O. Rädler

**Supplementary Data**

**S1 Supplementary Online Information Probabilistic Modeling**

**Probability distribution of active plasmids per cell**

The probability $P(X)$ of finding $X$ plasmids expressed in a given cell can be computed from a convolution of all underlying stochastic processes that occur prior to transcription initiation. Supposing $X$ plasmids have been activated, then $n \geq X$ plasmids first had to be delivered to the nucleus, with a probability $q$ for each plasmid to be expressed. This results in a binomial distribution with sample size $n$ and parameter $q$:

$$P(X|n) = \binom{n}{X} q^X (1-q)^{n-X} . \qquad [S1]$$

Two relevant stochastic processes determine the number of delivered plasmids $n$, namely, the number of complexes $C$ that arrive in the nucleus, and the number of plasmids in a given complex. We assume Poisson distributions for both, with means $\mu$ and $m$, respectively. Summing over all possibilities, we get the distribution

$$P(n) = \sum_{C=0}^{\infty} \frac{\mu^C}{C!} e^{-\mu} \sum_{n=0}^{\infty} \frac{(Cm)^n}{n!} e^{-Cm} \qquad [S2]$$

for $n$. Here we have used that the convolution of $C$ Poisson distributions, each with mean $m$, is again a Poissonian with mean $C \cdot m$.

Considering the previous two equations the overall probability of having $X$ active plasmids is

$$P(X) = \sum_{n=0}^{\infty} P(X|n) P(n) = \sum_{C=0}^{\infty} \frac{\mu^C}{C!} e^{-\mu} \sum_{n=0}^{\infty} \frac{(Cm)^n}{n!} e^{-Cm} \binom{n}{X} q^X (1-q)^{n-X} . \qquad [S3]$$

By interchanging the order of summation, shifting summation indices and using the normalization condition of the Poisson distribution, this can be rewritten as

$$P(X) = \frac{(mq)^X}{X!}e^{-\mu}\sum_{C=0}^{\infty}\frac{(\mu e^{-mq})^C}{C!}C^X .\qquad\text{[S4]}$$

Summing from $X = 1$ to infinity yields the transfection probability

$$TE := \text{Prop}(X > 0) = 1 - \exp\{\mu(e^{-mq} - 1)\}\qquad\text{[S5]}$$

which corresponds to Eq. **9** of the main text.

**Cotransfection Probabilities**

We are interested in the number of cells that are either monochromatic, dichromatic or not fluorescent at all. To compute the probabilities for each, a sum over all possible plasmid numbers $X$ has to be evaluated, with each term in the sum weighted with the probability of activation of zero, one, or two species, depending on the case being considered. If there are $i$ plasmids of one color in the nucleus, the probability that none are activated is $(1-q)^i$, while the probability that at least one is activated is $1-(1-q)^i$

The two cotransfection experiment setups were explained in the Material and Methods section. For uni-colored complexes (post-mixing), the total number of complexes can be subdivided into complexes of either color, yielding a binomial term in the complex number. Thus, for example, the probability of having non-fluorescent cells (not (CFP OR YFP)) is given by

$$\text{Prob}_{\text{post}}(\neg(C \vee Y)) = \sum_{C=0}^{\infty}\frac{\mu^C}{C!}e^{-\mu}\sum_{k=0}^{C}\binom{C}{k}\left(\frac{1}{2}\right)^C\left(\sum_{i=0}^{\infty}\frac{(km)^i}{i!}e^{-km}(1-q)^i\right)\left(\sum_{i=0}^{\infty}\frac{((C-k)m)^i}{i!}e^{-(C-k)m}(1-q)^i\right)$$
.**[S6]**

In the case of dual-colored complexes (pre-mixing), the total number of plasmids is binomial distributed between YFP and GFP, such that the probability of finding, for example, dichromatic cells (CFP AND YFP) is given by:

$$\text{Prob}_{\text{pre}}(C \wedge Y) = \sum_{C=0}^{\infty} \frac{\mu^C}{C!} e^{-\mu} \sum_{n=0}^{\infty} \frac{(Cm)^n}{n!} e^{-Cm} \sum_{i=0}^{n} \binom{n}{i} (1-(1-q)^i)(1-(1-q)^{n-i}). \quad [S7]$$

Similar expressions can be set up for all other cases. These can be algebraically simplified with the results given in Table SI.

From these expressions, it is easy to compute the cotransfection ratio,

$$r(\mu, m, q) = \frac{\text{Prob}(c \wedge y)}{\text{Prob}(c \vee y)} = \frac{\text{Prob}(c \wedge y)}{2 \cdot \text{Prob}(c \wedge \neg y) + \text{Prob}(c \wedge y)}. \quad [S8]$$

Figure S1 is a representative result for the cotransfection ratio, $r$ as a function of the transfection ratio, *TR,* for pre- and post-mixed complexes. Our model predicts that cotransfection is *enhanced* in pre-mixed complexes, and that the probability of cotransfection approaches 1 as *TR* approaches 100%. This is consistent with experimental results. The result shown in Figure S1 is particularly relevant in experiments, since one relies on cotransfection for the simultaneous delivery of two different plasmids.

**Distribution of Proteins**

The protein number distribution *P(G)* inherently carries the signature of the associated plasmid distribution *P(X)*. Ignoring intrinsic and extrinsic noise in gene expression the mean number of proteins can simply be computed from the distribution of plasmids Eq. **S4** and the expression factor:

$$\langle G \rangle = k_{\text{exp}} \langle X \rangle = k_{\text{exp}} \mu m q \quad [S9]$$

The mean protein number $\langle G \rangle$ can be obtained from single cell statistics. Additional relations are found between the parameters in Eq. **S9** by evaluating how the percentage of non-fluorescent cells, $p_0$ depends on them. $p_0$ is identical to the percentage of cells with no activated plasmids in Eq. **S4** or 1- *TR*, where *TR* is the transfection ratio.

$$p_0 := \text{Prob}(X = 0) = \exp\{\mu(e^{-mq} - 1)\}. \quad [S10]$$

Eliminating $\mu$ from Eqs. **S9** and **S10,** and with rearrangements, one finds

$$ae^a = we^w. \quad \text{[S11]}$$

where $a := \dfrac{\langle G \rangle}{k_{exp} \ln p_0}$ and $w := mq + a$. Solving Eq. **S11** for $w$ gives the Lambert W-function. Hence,

$$m_{eff} =: mq = \text{LambertW}(ae^a) - a \quad \text{[S12]}$$

which only depends on measurable quantities and $k_{exp}$. Fitting the expression factor as the only free parameter, $\mu$ and $m_{eff}$ can be determined from single cell data. Consequently, the distribution of active plasmids is set by Eq. **S4**. The distribution of proteins then follows by stretching the distribution of plasmids according to Eq. **6**. As argued in the main text, theory predicts discrete protein distributions, with peaks spaced by $k_{exp}$. Of course, there are additional noise sources like all post-transfectional fluctuations and limited measurement accuracy. To compare theory with experiment, we replaced the peaks of the discrete protein distribution by Gaussians with the same area and a standard deviation of 0.3 of each peak's position to approximate extrinsic noise. Figure 5 shows the complex, plasmid and protein distributions for a set of single cell data obtained from this theory. A full list of parameters for all four data sets is given in Table SII.

**Figure S1 (Schwake et al.)**

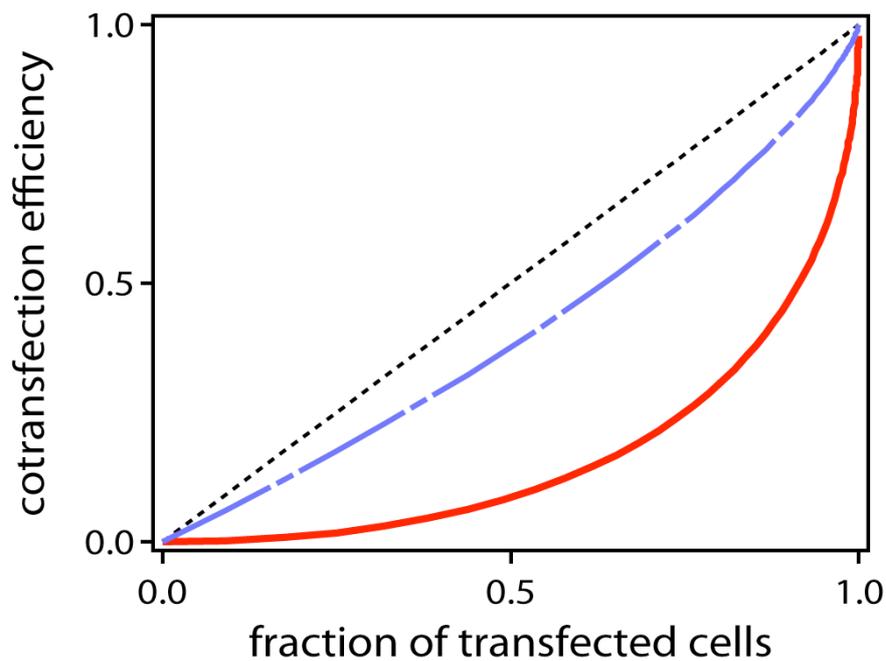

**Figure S1**. Theoretical prediction for the probability of cotransfection, *r,* as a function of the transfection ratio, *t*, in terms of percent cells transfected. The stochastic model of complex delivery predicts a strong discrepancy between pre-mixed and post-mixed (uni-colored) complexes. The analytic solution is given in the supporting information. Shown are parametric plots with *μ* varying between 0 and infinity, while $m_{eff}$ = 3.2 is held constant.

**Table SI (Schwake et al.)**

| | |
|---|---|
| $\text{Prob}_{\text{post}}(C \wedge \neg Y)$ | $\exp\left\{\frac{\mu}{2}(e^{-mq}-1)\right\} - \exp\left\{\mu(e^{-mq}-1)\right\}$ |
| $\text{Prob}_{\text{post}}(C \wedge Y)$ | $\left(1 - \exp\left\{\frac{\mu}{2}(e^{-mq}-1)\right\}\right)^2$ |
| $\text{Prob}_{\text{post}}(\neg(C \vee Y))$ | $\exp\left\{\mu(e^{-mq}-1)\right\}$ |
| $\text{Prob}_{\text{pre}}(C \wedge \neg Y)$ | $\exp\left\{\mu\left(e^{-\frac{m}{2}q}-1\right)\right\} - \exp\left\{\mu(e^{-mq}-1)\right\}$ |
| $\text{Prob}_{\text{pre}}(C \wedge Y)$ | $1 - 2\exp\left\{\mu\left(e^{-\frac{m}{2}q}-1\right)\right\} + \exp\left\{\mu(e^{-mq}-1)\right\}$ |
| $\text{Prob}_{\text{pre}}(\neg(C \vee Y))$ | $\exp\left\{\mu(e^{-mq}-1)\right\}$ |

**Table SI**. Probabilities of finding non-fluorescent, monochromatic, and dichromatic cells for pre-mixing (hetero-complex) and post-mixing (homo-complex) cotransfection.

**Table SII (Schwake et al.)**

| data | $p_0$ | $\langle G \rangle$ | $k_{exp}$ | $m_{eff}$ | $\mu$ | $\langle X \rangle^*$ |
|---|---|---|---|---|---|---|
| Synchronized cells, Lipofectamine | 0.70 | $1.55 \cdot 10^6$ | $1.3 \cdot 10^6$ | 3.21 | 0.37 | 4.0 |
| Asynchronous cells, Lipofectamine | 0.70 | $8.72 \cdot 10^5$ | $0.8 \cdot 10^6$ | 2.89 | 0.38 | 3.6 |
| Synchronized cells, PEI | 0.60 | $1.71 \cdot 10^6$ | $1.0 \cdot 10^6$ | 3.22 | 0.53 | 4.3 |
| Asynchronous cells, PEI | 0.77 | $6.06 \cdot 10^5$ | $0.7 \cdot 10^6$ | 3.18 | 0.27 | 3.8 |

**Table SII**. Parameters of the theoretical protein distribution as inferred from single cell data. The number of non-fluorescent cells, $p_0$, and the mean number of proteins, $\langle G \rangle$, have been determined experimentally. Fitting the expression factor $k_{exp}$, to give optimal agreement between theory and experiment, the mean number of delivered plasmids, $\mu$, and the mean number of activated plasmids per complex, $m_{eff} = m \cdot q$, are set by theory. From these parameters, the plasmid distribution, $P(X)$, is determined by Eq. **S4**. $\langle X \rangle^*$ is the mean number of expressing plasmids per fluorescent cell.

**S2 Supplementary online information to Materials and Methods**

**Materials.** Earle's MEM (Gibco, Catalog no. 41090-028), FBS (Invitrogen, Catalog no. 10106-185), Leibovitz's L-15 Medium (Gibco, Catalog no. 21083-027), Lipofectamine™2000 (Invitrogen, Catalog no. 11668-027) and OptiMEM (Gibco, Catalog no. 51985-026) were purchased from Invitrogen. 6-well culture plates (Falcon, Catalog no.353046) and thymidine (Catalog no. 6060-5) were purchased from VWR International GmbH. Linear PEI (25 kDa, Catalog no. 23966) was purchased from Polysciences, Europe GmbH, Eppelheim.

All plasmids were obtained from BD Biosciences: the pEGFP-N1 plasmid (Catalog no. 6085-1) consists of the transcriptional regulatory domain of cytomegalo virus (CMV) preceding the EGFP sequence; the pd2EGFP-N1 plasmid (Catalog no. 6009-1), derived from EGFP, contains a PEST amino acid sequence that targets the protein for degradation and results in rapid protein turnover; the pECFP-N1 plasmid (Catalog no. 6900-1) and the pEYFP-C1 plasmid (Catalog no. 6005-1), both derived from EGFP, encodes cyan and yellow fluorescent variants, respectively.

BD Living Colors EGFP Calibration Beads (Catalog no. 632394) were purchased from BD Biosciences. Sterile PBS and HEPES buffered saline (HBS) were prepared in-house. Trypsin-EDTA (Catalog no. Z-26-M) was purchased from c.c.pro GmbH.

**Cell culture.** A human bronchial epithelial cell line (BEAS-2B, ATCC) was grown in Earle's MEM supplemented with 10% FBS at 37°C in a humidified atmosphere, 5% $CO_2$ level. Cells were maintained at 85% confluence. Transfection was performed on both non-synchronized and synchronized cultures. A thymidine kinase double-block was performed to synchronize cells (Merrill 1998). Briefly, cells were cultured in synchronization medium (growth medium with 2mM thymidine) for 18 hours. Initial release was facilitated by an eight-hour incubation in regular growth medium. A second block was performed by incubating cells in synchronization medium for another 17 hours, which was followed by a three-hour incubation in regular growth medium to allow mid-S-Phase transfection.

**Transfection optimization.** Culture and transfection conditions were optimized at different levels. Different media on which the cells could grow were initially screened for autofluorescence and $CO_2$ independence. Leibovitz's L-15 Medium was found to be ideal for growing cells with minimal autofluorescence levels. Transfection was optimized by varying the Lipofectamine™2000:DNA and PEI:DNA ratios. Relative efficiencies were determined by Fluorolog Fluorescence Spectrofluorometer (Data not shown). Highest transfection efficiencies were obtained when 2μl of Lipofectamine™2000 or a Nitrogen:Phosphate ratio (N:P) = 8 of PEI:DNA was used with 1μg of DNA. EGFP expression stability was determined using time-course experiments; expression without photobleaching was found to be stable for at least four hours.

**Transfection.** BEAS-2B cells were grown to 80% confluence from an initial seeding density of 1 x $10^5$ cells/well in six-well plates at 37°C, 5% $CO_2$ level 24 hours before transfection. Cells were washed and the medium is replaced with 1 ml OptiMEM/well immediately before transfection. Optimized transfection procedures were performed using either Lipofectamine™2000 or PEI; 1 μg of pEGFP-N1 or pd2EGFP-N1, corresponding to 4·$10^{11}$ plasmids, is used for transfecting each batch of cells. For Lipofectamine-mediated transfection, separate 100 μl preparations of 1% w/v pEGFP-N1/OptiMEM or pd2EGFP-N1/OptiMEM and 2% v/v Lipofectamine™2000/OptiMEM were made at ambient temperature and allowed to stand for five minutes. The Lipofectamine transfection medium was prepared by adding the Lipofectamine solution to the plasmid solution. For PEI-mediated transfection, separate 100μl preparations of 1% w/v pEGFP-N1/HBS or pd2EGFP-N1/HBS and PEI (N:P = 8)/HBS were made and the solutions were combined by adding the PEI solution to the DNA solution. Transfection media were allowed to stand for an additional 20 minutes at ambient temperature to facilitate complex formation. Cells were incubated with 200 µl/well Lipofectamine or PEI transfection medium for 3 hours at 37°C, 5% $CO_2$ level. Negative controls were transfected with non-EGFP containing plasmids using the same delivery systems. After 3 hours of incubation the medium was removed, and cells were washed with PBS. Cells were reincubated with Leibovitz's L-15 Medium with 10% FBS prior to EGFP expression monitoring.

**Cotransfection.** For the cotransfection experiments, we choose ECFP and EYFP plasmids on account of their similarities in terms of size (4731bp vs. 4733bp), promoter (early promoter of CMV), DNA replication origin (SV40) and translational efficiency (Kozak consensus translation initiation site). These plasmids only differ in the position of the MCS-site (C-terminal in ECFP, N-terminal in EYFP), apart from the fluorescence gene that they express, which are necessarily different to distinguish ECFP- from EYFP- expressing cells. Cotransfection was performed with two kinds of preparations containing the same molar amount per plasmid. For one preparation, ECFP/Lipofectamine, EYFP/Lipofectamine, ECFP/PEI and EYFP/PEI were complexed separately. For the other, a mixture of ECFP and EYFP hetero-complexeswere complexed with Lipofectamine or PEI. Transfection using either the hetero-complexes (pre-mixed) or a mixture of homo-complexes (post-mixed) were performed as previously described. Cells were reincubated in growth medium and CFP and YFP expression was monitored by fluorescence microscopy after 24 hours.

**Instrumentation.** EGFP expression was monitored using a motorized inverted microscope (Axiovert100M, Zeiss) equipped with a temparature-controlled mounting frame for the microscope stage. SimplePCI (Compix) was used for all microscope controls. A mercury light source (HB 100) was used for illumination, and a reflector slider with three different filter blocks was used for detection. EGFP and d2EGFP were detected with filter set 41024 (Chroma Technology Corp., BP450-490, FT510, LP510-565), EYFP with F41-028 (AHF analysentechnik AG, BP500/20, FT515, BP535/30) and ECFP with F31-044 (AHF analysentechnik AG, BP436/20, FT455, BP480/40). An illumination shutter control was used to prevent bleaching.

**Data acquisition and quantitative image analysis.** Images were taken at 10× magnification, with a constant exposure time of 1s, at 10 minute-intervals for at least 30 hours post-transfection. Automated scanning and image capture for 25 view fields within a predefined 20mm diameter was programmed using the AIC module of SimplePCI. Briefly, the procedure captures individual brightfield images of the 25 view fields, then proceeds with a 30 hour loop where fluorescence images of each view field are taken at 10 minute intervals. At the end of the loop, brightfield images of the 25 view fields are taken again. Fluorescence images are consolidated into single image sequence files per view field. Negative control images were taken to assess lamp threshold values and autofluorescence, and were subtracted from corresponding image sequence files in SimplePCI to eliminate autofluorescence effects. To capture cell fluorescence over the entire sequence, regions of interest (ROIs) were manually defined around each cell in SimplePCI (Fig. 2). Changes in total gray measurements in individual ROIs were determined for each time point considered using a built-in function of SimplePCI.

**EGFP quantification.** EGFP is a Phe64Leu/Ser65Thr mutant of the *Auquorea victoria* protein, GFP (Heim et al. 1995). It exhibits an intense emission spectrum, higher photostability, and a half life of > 24 hours in mammalian cells (Bi et al. 2002); (Cotlet et al. 2001); (Tsien 1998). The accurate determination of the total number of EGFP molecules expressed in each cell is impeded by experimental limitations, notably bleaching and autofluorescence. We optimized culture, microscope and image acquisition settings to maximize the dynamic range while minimizing the effects of bleaching and autofluorescence. Our calibration procedure involved the use of an EGFP standard originally designed for flow cytometry. We used the microbead population coated with $1.16 \cdot 10^5$ EGFP molecules. For each transfection experiment, 50μl Bead4 of the EGFP-Calibration Kit was suspended in 1ml HBS in one well of a six-well plate. Fluorescence images of the beads were taken under the same conditions as the transfected cells. For each experiment, total gray values were obtained for at least 80 beads and an equal number of background regions adjacent to each bead. The average background corrected total gray value is obtained and used for converting intensity values to molecules of equivalent soluble fluorochrome (MESF) units. In our experiments, average calibration factors typically do not exceed a deviation of 11.2%. Calibrations between different assays have higher deviations due to illumination variations. Consequently, the calibration factor and the number of expressed EGFP molecules that we report are not absolute as a result of inherent and non-quantifiable differences in the quantum efficiencies of the EGFP chromophores inside the cells and the beads. However, relative intensity levels and the intensity distribution functions that are discussed in the paper are unaffected by the absolute calibration.

## References (Supplementary Online Information)